\documentclass[draft]{agujournal}
\draftfalse

\newcommand{\solphys}{{Solar Phys.}}

\newcommand{\pr}{{Phys. Rev.}}
\newcommand{\prl}{{Phys. Rev. Lett.}}

\newcommand{\pop}{{Phys. Plasmas}}
\newcommand{\ajp}{{Aust. J. Phys.}}
\newcommand{\apj}{{Astrophys. J.}}
\newcommand{\apjl}{{Astrophys. J. Lett.}}

\newcommand{\jgr}{{J. Geophys. Res.}}
\newcommand{\nat}{{Nature}}
\newcommand{\aap}{{Astron. Astrophys.}}

\newcommand{\ssr}{{Space Sci. Rev.}}
\newcommand{\planss}{{Planet. Space Sci.}}

\newcommand{\mnras}{{Mon. Not. Roy. Astron. Soc.}}

\newcommand{\aapr}{   {Astron. Astrophys. Rev.}}

\newcommand{\pasj}{ {Publ. Astron. Soc. Japan}}
\newcommand{\pasa}{ {Publ. Astron. Soc. Aust.}}

\journalname{JGR-Space Physics}

\begin{document}

%
%


\title{Current-driven flare and CME models}

%
%




\authors{D.B. Melrose}


\affiliation{\null}{
SIfA, School of Physics, University of Sydney, NSW 2006, Australia\\
ORCiD: 0000-0002-6127-4545}




\correspondingauthor{Don Melrose}{donald.melrose@sydney.edu.au}




\begin{keypoints}
\item Magnetic energy release is due to reconfiguration of large-scale currents
\item Energetically important coronal currents are unneutralized
\item A flare-associated current is driven by the electromotive force
\end{keypoints}

%
%


\begin{abstract}
Roles played by the currents in the impulsive phase of a solar flare and in a coronal mass ejection (CME) are reviewed. Solar flares are magnetic explosions: magnetic energy stored in unneutralized currents in coronal loops is released into energetic electrons in the impulsive phase and into mass motion in a CME.  The energy release is due to a change in current configuration effectively reducing the net current path. A flare is driven by the electromotive force (EMF) due to the changing magnetic flux. The EMF drives a flare-associated current whose cross-field closure is achieved by redirection along field lines to the chromosphere and back.  The essential roles that currents play are obscured in the  ``standard'' model and are described incorrectly in circuit models.  A semi-quantitative treatment of the energy and the EMF is provided by a multi-current model, in which the currents are constant and the change in the current paths is described by time-dependent inductances. There is no self-consistent model that includes the intrinsic time dependence, the EMF, the flare-associated current and the internal energy transport during a flare. The current, through magnetic helicity, plays an important role in a CME, with twist converted into writhe allowing the kink instability plus reconnection to lead to a new closed loop, and with the current-current force accelerating the CME through the torus instability.
\end{abstract}

%
%

%


%
%
%
%

\section{Introduction}
\label{sect:Introduction}

Solar flares are magnetic explosions: magnetic energy builds up through currents stored in the corona above an active region over days to weeks, and is partly converted into kinetic energy over $10^2$--$10^3\,$s during a flare. This review is concerned with the roles these currents play in the impulsive phase of a flare, and also in coronal mass ejections (CMEs) that are associated with some flares. These roles are obscured or treated incorrectly in many older models for flares. Specifically, the currents are obscured in two-dimensional (2D) versions of the ``standard'' model, and the currents are treated incorrectly in most circuit models. Such models are also stationary, in the sense that they include no intrinsic time dependence, whereas an essential ingredient in a magnetic explosion is the time-changing magnetic field. The only effective tool available to describe time-dependent global electrodynamics is the integrated form of Maxwell's equations. The time-dependent magnetic field implies an electromotive force (EMF) (The integrated form of Faraday's equation implies that for any closed path the EMF, $\Phi$, which is the line-integral of the electric field around the closed path, is equal to minus the rate of change of the magnetic flux enclosed by the path.) that drives a flare-associated current, against which it does the work that results in the magnetic energy release. However, these essential ingredients, specifically the EMF and the flare-associated current, are either implicit or missing in most flare models. An intrinsically time-dependent description of the electrodynamics must involve circuit-like ingredients, without the misleading assumptions made in most circuit models. Currents play a different role in a CME, whose acceleration can plausibly be attributed to a current-current force.

The flare energy output \citep[e.g.,][]{Sv76,T-HE88,S92,A04} ranges, from one flare to another, over many orders of magnitude. Much emphasis is placed on the most energetic flares (``big-flare syndrome''), rather than the much more frequent, less energetic flares. Flares are classified  A, B, C, M or X (e.g., according to the peak energy flux in 100 to 800 picometer X-rays measured on the GOES spacecraft near the Earth, with class A $<10^{-7}\rm\,W\,m^{-2}$ and class X $>10^{-4}\rm\,W\,m^{-2}$), with the rate of occurrence of flares decreasing rapidly with increasing energy output. The released magnetic energy appears as kinetic energy, in energetic particles,  mass motions and heat. All flares have an impulsive phase, in which the energy released goes predominantly into ``bulk energization'' of a copious number of 10--$20\,$keV electrons, which produces the main signatures of a flare: hard X-ray bursts (HXBs), H$\alpha$ emission (and white light in some energetic flares) by precipitating electrons, and type III radio bursts by escaping electrons. The area that brightens in H$\alpha$ correlates with the energy released, and underlies the reference to flare class in terms of size, with A being smallest and X being largest. There is a long-standing ``number problem'' \citep{B71,B76} in that many more electrons precipitate, up to about $10^{39}$ in a large flare, than were present in the coronal flux loop prior to the flare. Moreover, the rate of electron precipitation, ${\dot N}$ of order $10^{36}\rm\,s^{-1}$, would imply an impossibly large current, $e{\dot N}$, of order $10^{17}\rm\,A$, requiring a ``return current''  to provide current neutralization \citep[e.g.,][]{vdO90}. The number problem and the return current problem need to be addressed in a realistic model for the energy release in the impulsive phase. For illustrative purposes, when discussing semi-quantitative aspects of the impulsive phase of a flare, the following numbers are chosen, reflecting a moderately large flare: total energy $10^{24}\,$J, power $10^{21}\,$W and rate of precipitation ${\dot N}=10^{36}\rm\,s^{-1}$ of electrons with energy $\varepsilon=10\,$keV $\approx10^{-15}\,$J. The current, which has long been known from vector magnetograms \citep{AC67}, is taken to be $I=10^{11}\,$A, and the EMF to be $\Phi=10^{10}\,$V, such that the power is $I\Phi=\varepsilon{\dot N}$. 

A flare with an associated  coronal mass ejection (CME ) is said to be ``eruptive''. In an eruptive flare the energy goes predominantly into  the CME;  X-ray jets \citep{Setal92} are another form of mass motion. CMEs and flares are associated, but not all flares are eruptive. The statistical probability of a CME increases from C class to X class \citep[e.g.,][]{CRZ17}, but is not unity even for the largest flares. Which of flare and CME is cause and which is effect has long been (and remains) controversial. The acceleration of a CME can be attributed to a current-current force, which can be derived from the Lorentz force in an approach based on magnetohydrodynamics (MHD), whereas the EMF and bulk energization of electrons, due to a parallel electric field $E_\parallel$ \citep{H85}, require a non-fluid-based approach. 

In brief, a flare model needs to explain the energy release into energetic electrons in the impulsive phase, and it needs to allow but not to require the acceleration of a CME or a jet. Currents play a central role in both these aspects of a magnetic explosion.

General properties of currents related to flares are discussed in Section~\ref{sect:currents}, including some historical remarks on how relevant ideas evolved. The distinction between neutralized and unneutralized currents is emphasized. The ``standard'' model and circuit models for flares are discussed critically in Section~\ref{sect:circuit}. The description of magnetic energy in terms of currents, and how the energy changes during a flare is discussed in Section~\ref{sect:energy}. The role of magnetic helicity in flares and CMEs is discussed in Section~\ref{sect:helicity}. The results are discussed in Section~\ref{sect:discussion}.

\begin{figure}
    \begin{center}
 \includegraphics[width=0.8\hsize]{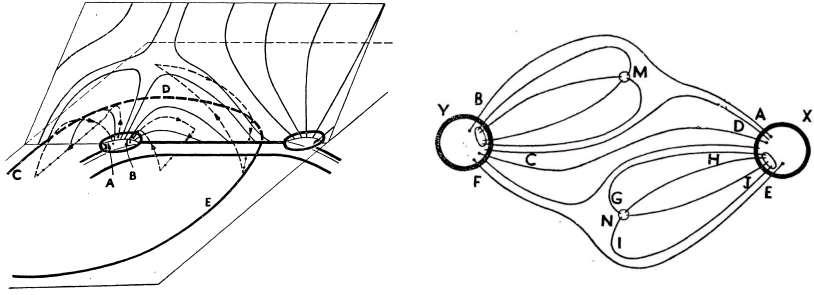}
\caption{Two cartoons from \citet{G48}; his Fig.~1, on the left, illustrates the magnetic field lines in a plane that intersects two sunspots (indicated by dark rings) with currents indicated by broken lines; his Fig.~4, on the right, indicates the magnetic connectivity between a pair of large spots, X and Y (large dark circles), and two small spots, M and N (small light circles). } 
\label{fig:G48}
\end{center}
   \end{figure}

 \section{Currents in the solar corona}
 \label{sect:currents}
 
In this section I first comment on an early flare model, and how subsequent ideas relating to flares then developed, emphasizing the role of currents. I then introduce a classification of currents, identifying the currents that are energetically relevant to flares.

\subsection{The role of currents in early flare models}

An early model for ``chromospheric flares'' was proposed by \citet{G46,G47,G48}. Two of the cartoons from \citet{G48} are reproduced in Figure~\ref{fig:G48}.  Based on his Fig.~1, Giovanelli identified ``two types of current system'', one ``in a sheet towards and through the neutral point'' and the other ``into and out from a sunspot'', and in his Fig.~2 (not shown), he indicated the possible locations of neutral points. His Fig.~4 shows a quadrupolar structure in which he identified current paths along ``AB, CD, EF, joining the major sunspots.''  There is a remarkable similarity between Giovanelli's ideas and modern ideas on magnetic structures and magnetic energy release, which is intrinsically three dimensional (3D) \citep{L05,P11,Petal15}.  Furthermore, \citet{G48} discussed the inductive electric field explicitly and estimated its value, from which one can infer that the EMF is of order $10^9\,$V, similar to an earlier estimate of $10^{10}\,$V by \citet{S33} in a related context. 

Giovanelli's model motivated the development of ideas for magnetic reconnection by \citet{D53,D58}, leading to the Sweet-Parker \citep{S58,P57}, Petschek \citep{P64} and subsequent models for magnetic reconnection. These reconnection models are 2D, and the original version of the ``standard'' or CSHKP flare model  \citep{C64,S66,H74,KP76} based on them is also 2D.  The current in any 2D model for a magnetic field is perpendicular to the 2D plane, precluding any discussion of current closure within a 2D model.

 \subsection{Classification of currents}

It is helpful to distinguish transient currents and slowly-varying currents, depending on whether the time scale on which they change is less than or much greater than the timescale for a flare, and to separate slowly-varying currents  that may be relevant to solar flares into three classes:\\
 \noindent
1) Currents that close below the photosphere: these currents produce only a potential magnetic field in the corona, and this potential component does not change during a flare  because the timescale for currents  below the photosphere to change is much longer than the timescale for a flare.\\
 \noindent
2) Currents confined entirely to the corona:  by hypothesis such currents must close across field lines in the corona, requiring that the Lorentz force be balanced by a pressure force (or by inertia for a transient current), which is assumed negligible in ``force-free'' models for the coronal magnetic field; such currents cannot be significant energetically before or after a flare.\\
 \noindent
3) Currents that flow in coronal magnetic loops, flowing into the corona through the photosphere at one footpoint of the loop and flowing back through the photosphere at the other footpoint: such currents provide the magnetic free energy available to power a flare. \\

Currents in the third class can be further separated into two subclasses:\\
 \noindent
3a) Neutralized currents, in which there is a direct current and a return current flowing along the coronal loop, with the direct and return current closing across field lines at or below the photosphere. Such currents can build up after the flux loop has emerged, due to twisting or shearing motions. \\
 \noindent
3b) Unneutralized currents, in which there is no return current in the corona, with the current closing deep inside the Sun.\\

 \noindent
The distinction between neutralized and unneutralized currents is important in distinguishing between flare models.  On the one hand, a neutralized current is implicit in circuit models in which the driver is a photospheric dynamo which turns on at the start of the flare. On the other hand, the appearance of a new flux loop, as it emerges through the photosphere, suggests that it is already twisted \citep{Letal96}, implying that an unneutralized current is already flowing. How the current evolves in response to photospheric motions after the flux has emerged is still being debated \citep{Tetal14,Detal15}.

\subsection{Are currents neutralized or unneutralized?}

This question has been discussed recently from both observational \citep{Getal12} and theoretical \citep{Tetal14} viewpoints, but nevertheless remains somewhat controversial. The original controversy \citep{M95,M96,P96} arose partly from the then new evidence that emerging flux tubes are twisted \citep{Letal96}, indicating that the (unneutralized) current is flowing in the flux tube before it emerges, rather than developing as a result of twisting or shearing after the flux tube has emerged. The observational question \citep{Getal12} concerns whether vector magnetograms show that the upward and downward  currents injected into a bipolar region are balanced separately on either side of the magnetic neutral line (neutralized) or only when integrated over a larger area that includes both bipoles (unneutralized). The theoretical question is whether the emergence of a flux tube that is neutralized below the photosphere leads to neutralized currents in the corona. This question has been explored through MHD simulations \citep{TK03,ADG05,Tetal14,Detal15}, leading to the conclusion that the resulting coronal currents are substantially unneutralized when there is shear across the polarity inversion line. From a physical viewpoint, the answer to this question is important when considering both the free energy that is available in a flare, and the force that drives this energy release. In this paper, relevant currents are assumed to be unneutralized, but before making this assumption, it is relevant to comment further on implications of making the opposite assumption, that coronal currents are neutralized.

\subsection{Neutralized currents}

Two unrelated assumptions that would imply a neutralized coronal current are, first, that a coronal flux tube can be treated as an isolated entity and, second, that a coronal current is driven by a photospheric motion (a ``photospheric dynamo'').

Suppose that a current-carrying magnetic flux tube can be regarded as an isolated entity. In the idealized case of a cylindrical, force-free, current-carrying flux tube of radius $R$, an unneutralized current implies an azimuthal component, $B_\phi=\mu_0I/2\pi r$ at $r>R$. This field outside the flux tube would lead to an interaction with any neighboring current-carrying flux tube through the Lorentz force ${\bf J}\times{\bf B}$, where ${\bf J}$ is the current density in the neighboring flux tube and ${\bf B}$ is identified with the field $B_\phi$; this violates the assumption that the flux tube is isolated. Current-neutralization is required to ensure $B_\phi=0$ at $r>R$ so that the flux tube is isolated from its neighbors. For any given radial current profile within the flux tube, it is always possible to construct a force-free surface current layer, at $r\lesssim R$, to satisfy the neutralization condition $B_\phi=0$ at $r>R$ \citep{MNB94}.  For the purpose of further discussion, suppose the return current is confined to the surface, $r=R$,  and compare the energy stored in a neutralized flux tube and an unneutralized flux tube with the same internal current profile and no return surface current. The magnetic energy stored inside of the flux tube, at $r<R$, is the same in both cases, and is determined by the integral of $B_\phi^2/2\mu_0$ over the volume of the flux tube. This internal magnetic energy is free energy, in the sense that it can be reduced to zero by a change in the current profile, such that the direct current becomes a surface current at $r=R$. In the neutralized case this corresponds to the direct and return currents cancelling, so that one has $B_\phi=0$ at  $r>R$. In the unneutralized case,  a change in the current profile at $r<R$ does not affect $B_\phi$ at $r>R$. In the latter case, for a cylindrical model, the magnetic energy per unit length in the field $B_\phi$ at $r>R$ diverges logarithmically. Of course, this divergence is artificial and the magnetic energy associated with any realistic model for a closed current is finite.  Although the cylindrical case is obviously over-idealized when discussing the energy, it does indicate how magnetic ``free'' energy can be stored in unneutralized currents. The current in a given flux tube implies $B_\phi$ at $r>R$ and the associated magnetic energy can change if the current is redistributed between flux tubes during a flare. In this idealized model, the axial magnetic field corresponds to the potential field due to subphotospheric currents which cannot change significantly during a flare. A semi-quantitative treatment of magnetic energy storage in unneutralized current systems is given by the multi-current model in section~\ref{sect:energy}.

Neutralized currents arise in models in which the coronal currents are assumed to be driven by a photospheric dynamo, that is, currents attributed to photospheric motions that twist or shear the magnetic field, cf.\ the circuit model Figure~\ref{fig:Spicer}. Consider a flux loop that initially has no current, and is subjected to a twisting motions: a torsional fluid motion is imposed (by some unspecified torque) at an ``active'' footpoint. The driving torque is opposed by a torque due to the Lorentz force, ${\bf J}\times{\bf B}$, where the cross-field current is driven as a reaction to the imposed torque. If the torque is imposed suddenly, the resulting cross-field current propagates in an Alfv\'enic front  \citep{M92} to the other ``passive'' footpoint where it is reflected, with the reflection coefficient depending on the effective resistance of the reflection layer. Over many Alfv\'en propagation times, this resistance damps the Alfv\'enic front,  resulting in a closed current loop that has direct and return paths along coronal field lines, closing across field lines at the active and passive footpoints due to the cross-field (Pedersen) conductivity of the reflection layer. In  an idealized cylindrical model for a twisted flux loop, there is a direct current inside the flux loop, and a return current on its surface; in a more realistic cylindrical model, the direct and return current are both distributed in $r$ such that the axial current reverses sign at some $r$ \citep[e.g.,][]{MNB94}. In the case of a shearing motion, similarly imposed on an arcade of flux loops, the direct and return currents  can be on field lines either on the inside or the outside of the arcade \citep[e.g.,][]{Detal15}. 

Whether or not photospheric flows do twist and shear the coronal magnetic field and set up such a neutralized current is uncertain. An alternative interpretation is that twisting and shearing motions are only apparent, reflecting a rising twisted or sheared magnetic field \citep{Letal96}.

\subsection{Return current}

There is one flare-related context where current neutralization is required: a return current is needed to neutralize the direct current associated with precipitating electrons. The implied current $\approx10^{17}\rm\,A$ greatly exceeds the maximum current in a coronal flux loop \citep{H85}, requiring a return current such that the net current is consistent with a redirected coronal current ($\approx10^{11}\rm\,A$). Early models for the return current \citep{BM77,BB84,SS84,vdO90} assumed that it is co-spatial with the direct current. In these models, the precipitating electron beam was assumed to be turned on in the corona, and that the return current was attributed to ambient electrons accelerated by either an electrostatic or inductive electric field \citep{vdO90} in response to the postulated beam of electrons. Such a model cannot apply if the precipitating electrons are accelerated by a parallel electric field: the same parallel electric field cannot accelerate the precipitating electrons downward and the electrons in the return current upward on the same field line \citep{H85,EH95}, cf.\ also \citet{FH08}. 

An unrelated criticism of return-current models is that (in the equations used) the motion of the ions is neglected, thereby precluding any role for Alfv\'en waves. When the ions are included, inductive effects propagate at the Alfv\'en speed. A plausible model is that, as the flare develops, the direct and return currents build up together, forming a closed current loop, with direct and return parallel currents on neighboring field lines and closure across field lines at two end points. Such a current system is familiar in standing-Alfv\'en-wave models for auroral arcs \citep{A70,SH73,Metal77,MC78,K96}. This analogy with the acceleration of auroral electrons suggests that the acceleration of electrons occurs due to a parallel electric field in the upward current region \citep[e.g.,][]{Eetal02}.  In the application to a flare, the model requires a large number of neigboring up- and down-current paths \citep{H85,MW14}. Specifically, with the numbers adopted here, $10^6$ up and down currents of $10^{11}\rm\,A$, each corresponding to $10^{30}$ electrons per second, are required to give a net $10^{36}\rm\,s^{-1}$ precipitating electrons on the up-current paths. The down-current paths are due to chromospheric electrons drawn up into the corona. Such a model solves the ``number problem'' with the source of the electrons being the chromosphere, with the electrons precipitating at one footpoint probably drawn from the chromosphere at the conjugate footpoint  \citep{EH95}.

The direct current due to precipitating electrons and the return current that neutralizes it are transient, in the sense that they are present only during the impulsive phase. They are not relevant to magnetic energy storage in the corona, and may be regarded as part of the flare-associated current that flows in response to the EMF during a flare. 

\subsection{Current-current interactions}

An essential ingredient in a flare model is the driver of the energy release; in a magnetic explosion the driver may be interpreted either as the EMF or as the Lorentz force. The equivalence of these two drivers is discussed below; here the interpretation in terms of the Lorentz force is adopted for the purpose of discussion. On a global scale this driver may also be interpreted as the current-current force between current-carrying flux loops.  Consider two loops labeled 1 and 2. Let the magnetic field due to the current in loop~1 be  ${\bf B}_{12}$ at loop 2, resulting in a Lorentz force ${\bf J}_2\times{\bf B}_{12}$ on loop 2. Similarly, the Lorentz force on loop 1 is ${\bf J}_1\times{\bf B}_{21}$, where the roles of 1 and 2 are interchanged. Such a current-current force is only present if the currents are unneutralized. If the current in either loop is neutralized, the magnetic field due to that current is zero (by hypothesis) outside the loop. 


The qualitative effects of the current-current force operating between any two unneutralized currents can be understood by noting that like currents attract and unlike currents repel. Such forces were invoked for various purposes in the early literature on solar magnetic structures. One example is the \citet{GH60} model for a  flare; this model involves two magnetic loops with like currents and  oppositely directed axial magnetic standard model for sol
: the current-current force draws the two together and the energy release is attributed to annihilation of the axial fields. Another example is a model for the support of a filament against gravity due to the repulsion between the axial current in the filament and an oppositely directed current in a mirror image below the photosphere \citep{KR74,vTK78}. A related idea \citep{A78} is a model for acceleration of a CME involving a current around a closed path, with the lower part of the path fixed in the photosphere: the force on the upper part drives the CME outwards. 

More recently, the torus instability \citep{Sh66,B78} has been invoked in connection with the acceleration of a CME \citep{KT06,Zetal14,Zetal15,Metal17}, cf.\ the detailed analysis by \citet{DA10}. This is similar to the model of \citet{A78} in that the acceleration is driven by the nonlocal, current-current force between unneutralized currents on opposite sides of the torus. Such driving by the current-current force applies only if the currents are unneutralized.

\subsection{Flare-associated current}

The EMF around any closed path tends to drive a current around this closed path. This current is referred to here as the flare-associated current. Currents cannot flow around most closed paths in the corona, and the EMF is relevant only for the path around which the current actually flows, and this is not known.  Currents can flow freely only along field lines, and relevant field lines do not form closed paths in the corona. This suggests that the current path is along one set of field lines and back along another set of field lines, with cross-field current closure at two end points. However, cross-field currents can flow only under special circumstance in the corona, in response to an imposed force, and it is at best uncertain whether any significant current closure occurs in the corona. An alternative, assumed here, is that the end points are in the chromosphere, where cross-field current can flow and provide closure due to the Pedersen conductivity of the partially ionized plasma there.

The flare-associated current is an important concept that has been given little attention in the literature.  Several further comments on it are appropriate. First, the net effect of the flare-associated current over the duration of a flare must be to change the initial current configuration into the final current configuration. Second, the current remains approximately field aligned, over most of its path, during a flare implying that the  coronal current pattern and the coronal magnetic field pattern evolve together. Third,  assuming that the actual cross-field current closure is in the chromosphere, a plausible model for the current path is analogous to the current-wedge model for a geomagnetic substorm \citep{MRA73}.   Such a model requires that the magnetic energy released in the corona get converted into an Alfv\'enic flux that transports it to the acceleration region near the chromosphere where it is transferred to energetic electrons \citep{H12,M12a,M12b,MW13,MW14}. Detailed magnetospheric models for the generation of the Alfv\'enic flux  \citep{BH96,Betal99,V05} may provide a helpful guide as to how the released magnetic energy is converted into Alfv\'enic form in the solar-flare analog.

As remarked above, the flare driver can be interpreted either as the EMF or the Lorentz force, and the equivalence of these follows from the final two of the foregoing points. Locally, the rate magnetic energy is released per unit volume is the rate the Lorentz force does work, that is, ${\bf v}\cdot({\bf J}\times{\bf B})={\bf J}\cdot{\bf E}$, where ${\bf v}$  is the plasma flow velocity and ${\bf E}=-{\bf v}\times{\bf B}$ is the convective electric field. The localization of the EMF across regions of cross-field current flow implies that it can be identified with the line integral of $-{\bf v}\times{\bf B}$ in these regions. Thus one may regard the driver as either the Lorentz force doing work on the fluid, or as the EMF doing work against the current, with these being equivalent in ideal MHD. 

\section{Critique of standard and circuit models}  
\label{sect:circuit}

The standard model and circuit models for solar flares are misleading in relation to the role of currents, due either to the currents being ignored or treated incorrectly.

\begin{figure}
\begin{center}
 \includegraphics[width=0.7\hsize]{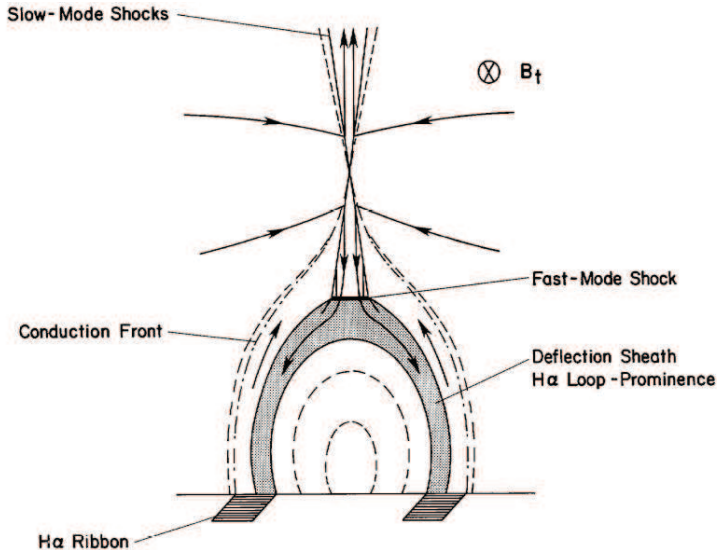}
 \caption{A specific example of the standard model for a solar flare: reconnection is assumed to occur on open field lines above a closed loop, with the reconnected field lines transporting energy downward, accelerating electrons at a shock where the flow meets the closed loop, and upward in a CME  \citep{FM86}.}
 \label{fig:standard}
 \end{center}
 \end{figure}

\subsection{Standard or CSHKP model}

The standard model for solar flares, \citep[e.g.,][]{PF02,Shibata05}, was developed primarily to explain eruptive flares.  An example of the many cartoons used to describe the standard model for a solar flare is shown in Figure~\ref{fig:standard}.  The reconnection leads to plasma outflow from the reconnection site, with the upward flow assumed to result in a CME \citep{S66} and the downward flow assumed to result ultimately in the various other signatures of a flare. There are two assumptions in the standard model that lead to the role of the current being obscured: the 2D assumption and an implicit steady-state assumption. 

In 2D versions, such as the one illustrated, the current is perpendicular to its page, precluding any discussion of current closure. Although it has long been recognized that reconnection is intrinsically 3D \citep{G88,LF91}, the CSHKP model is still widely regarded as a ``standard''  model for solar flares. Only recently has the current been discussed explicitly in 3D versions of the standard model \citep[e.g.,][]{Jetal14}. 

From an electrodynamic viewpoint, a serious weakness in the standard model is that there is no time dependence, and hence no EMF. As a consequence, the actual driver of the flare is replaced by a proxy: steady-state inflows and outflows at appropriate boundaries, with nothing actually changing with time in the model itself. Not only the EMF but also the flare-associated current are excluded by the steady-state assumption in the standard model. It is essential that the current change, in order for magnetic energy to be released, and this ingredient is also either missing or is implicit in the boundary conditions.  There are some notable exceptions where the time-dependence and the EMF are included in eruptive flare models that indicate how the energy is transferred to the motion of a CME \citep[e.g.,][]{MK89,LF00,JAD15}. Nevertheless, how the required enormous EMF, of order $10^{10}\,$V, is to be included is ignored in most flare models.

\subsection{Circuit models}

\begin{figure}
\begin{center}
 \includegraphics[width=0.8\hsize]{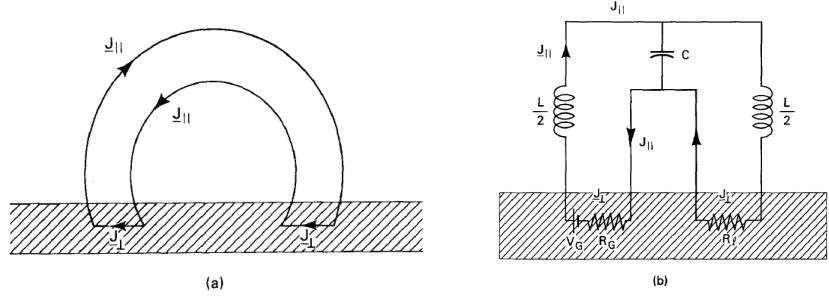}
  \caption{The conventional circuit model for a solar flare \citep{S82} discussed critically in the text.}
  \label{fig:Spicer}
  \end{center}
 \end{figure}

Consider the circuit model in Figure~\ref{fig:Spicer}. Such models are misleading in several ways. First, as the figure on the left indicates, the current is  assumed to have both a direct path and a return path through the corona, implying that this is a neutralized current.  This current is assumed to close across field lines in the shaded region, which represents an assumed resistive layer at the photosphere. Such cross-field closure implies a Lorentz force that must be balanced by another force or by inertia in the photosphere. The model does not include a pressure force to balance this Lorentz force.  If the Lorentz force is balanced by inertia, the cross-field current propagates away Alfv\'enically, implying that this current can be maintained at the photosphere only for an Alfv\'en propagation time, which is shorter than the days to weeks over which the stored magnetic energy builds up \citep{WM94}.  The model is driven by a photospheric dynamo, indicated by the voltage source at one ``active'' footpoint in the circuit diagram on the right of Figure~\ref{fig:Spicer}.  However, the assumption that the flare is powered by a photospheric dynamo  is inconsistent with the flare being powered by release of magnetic energy stored in the corona.  A further feature of the model is that the energy release is attributed to the turning on of a resistance, denoted by $R(t)$, and the modeling of magnetic energy release in a flare in terms of a resistance in a circuit suggests that the power released goes into heat, whereas it goes into energetic electrons in the impulsive phase and into mass motion in an eruptive flare.

The sudden turning on of a resistance in the corona was referred to as ``current interruption'' by \citet{AC67}, who assumed the resistance to be due to a double layer. However, the sudden turning on of a region of enhanced resistance, $R_c$, in a current-carrying coronal loop causes the current to be partly redirected around the the resistive region  \citep{M95,P96}. Specifically, the turning on sets up a cross-field current and electric field that propagate away in Alfv\'enic fronts,  which reflect from the footpoints and dissipate only over many Alfv\'en propagation times due to the assumed resistance at the reflection points \citep{M92}. In a cylindrical model, the fraction of the initial current that is deflected around the resistance is $R_c/(R_A+R_c)$, with $R_A=\mu_0v_A/4\pi$. The maximum dissipation occurs for $R_c=R_A$, and this energy is dissipated in the photosphere by the assumed resistance at the reflection points. 

In summary, criticisms of the circuit model in Figure~\ref{fig:Spicer} include: the EMF is not treated correctly, being replaced by a photospheric dynamo, the current path is mis-identified, and the location and nature of the dissipation are mis-interpreted. Such criticisms have not only led to specific circuit models being viewed as unacceptable, but also to avoidance of circuit concepts altogether in discussing flare electrodynamics. However, some circuit concepts are essential ingredients in understanding the global electrodynamics. Criticisms of the circuit model in Figure~\ref{fig:Spicer} do not invalidate the concept that a flare can be interpreted in terms of an EMF driving a flare-associated current around a closed circuit. 

\section{Magnetic energy in terms of currents}
\label{sect:energy}

An aspect of circuit models that needs to be retained in describing the electrodynamics of flares is the description of magnetic energy in terms of currents. 

\subsection{Magnetic energy in terms of currents}

The total magnetic energy (in all space) due to a current ${\bf J}({\bf x})$ confined to a volume $V$ is  \citep{J75}
\begin{equation}
E_{\rm mag}={1\over2}\int_V d^3{\bf x}\,{\bf J}({\bf x})\cdot{\bf A}({\bf x}),
\qquad
{\bf A}({\bf x})=\frac{\mu_0}{4\pi}\int_V d^3{\bf x}'\,
{{\bf J}({\bf x}')\over|{\bf x}-{\bf x}'|},
\label{Edef}
\end{equation}
giving
\begin{equation}
E_{\rm mag}={\mu_0\over8\pi}\int_V d^3{\bf x}\int_Vd^3{\bf x}'\,
{{\bf J}({\bf x})\cdot{\bf J}({\bf x}')\over|{\bf x}-{\bf x}'|},
\label{Emag1}
\end{equation} 
which may also be written as the energy density $|{\bf B}|^2/2\mu_0$ integrated over the volume $V$ plus a surface integral that is interpreted as the magnetic energy outside $V$ due to the current inside $V$.  The volume  may be separated into that inside the Sun plus that of the corona, $V=V_{\rm in}+V_{\rm cor}$, but the double integral does not allow one to ignore the contribution from $V_{\rm in}$ by assuming it to be a constant. One way of overcoming this difficulty is to replace $V_{\rm in}$ by a mirror image of $V_{\rm cor}$ \citep[e.g.,][]{KR74}.  Such a model includes a (repulsive) current-current force between the coronal portion of the flux loop and its mirror image, and this force was invoked by \citet{KR74} to support a prominence against gravity.
\citet{M95} and \citet{HMH98} used such a model to estimate the energy change during a flare, with each coronal loop being modeled as a semi-torus.  


In a multiple-current model, the continuous current in the corona is modeled as a sum of discrete currents. The total volume over which spatial integrals in equation (\ref{Emag1}) are performed is separated into a sum of discrete volumes each corresponding to the product of the integrals perpendicular and parallel to the local current path. For the $i$th discrete current, the integral over the cross section perpendicular to ${\bf J}$ gives the current, $I_i$, and the cross section is assumed to adjust along the current path such that $I_i$ is constant. For a set of discrete currents with $i=1,2,\ldots$, this allows equation (\ref{Ecm}) to be written as
\begin{equation}
E_{\rm mag}={\textstyle\frac{1}{2}}\sum_{ij}M_{ij}I_iI_j={\textstyle\frac{1}{2}}\sum_iL_iI_i^2+\sum_{i<j}M_{ij}I_iI_j,
\label{Ecm}
\end{equation}
where the $M_{ij}$ depend only on the geometry, defined by the way the total current is split up into discrete currents. The $L_i=M_{ii}$ are self-inductances, and the $M_{ij}$ for $i\ne j$ are mutual inductances. 

In the solar corona, the force-free condition implies $\mu_0{\bf J}=\alpha{\bf B}$ with $\alpha$ constant along magnetic field lines. The discrete volumes then correspond to discrete magnetic flux tubes. For the $i$th discrete volume one may write $\mu_0I_i=\alpha_i\Psi_i$, where $\Psi_i$ is the magnetic flux in the $i$th flux tube. A force-free magnetic field is said to be linear if all the $\alpha_i$ are the same and nonlinear is the $\alpha_i$ are different. The (nonlinear) force-free assumption allows the magnetic and current pattern in the corona to be calculated given the boundary conditions at the solar surface \citep{WS12,Betal16}. For any such a model, the magnetic energy may be written in the form (\ref{Ecm}).

\subsection{Single-loop model}

During a flare, there is insufficient time for the magnetic field and the current inside the Sun (below the photosphere) to change significantly. The change in the current pattern in the corona may be described in terms of changes in the paths along which the discrete currents flow, that is, in terms of the geometry that determines the $L_i$ and $M_{ij}$, and how this geometry changes during a flare. (Such changes require a change in the magnetic/current configuration at the footpoints of the flux loops; these changes occur in only a small fraction of the magnetic flux tube and are assumed to be unimportant in the overall flare energy budget.)  The simplest example of magnetic energy release due to a change in the geometry without any change in the magnitude of the current is the shrinkage of a single current-carrying flux loop \citep{FA96,FH08,H16}. This corresponds to equation~(\ref{Ecm}) with only one current, $I_i$. With $L_i$ proportional to the length of the current path, the stored magnetic energy is reduced due to the shortening of the length of the loop as it  shrinks.  However, it should be emphasized that any circuit model based on isolated flux tubes neglects essential physics (notably the interaction with neighboring flux tubes) that can only be included through detailed MHD modeling, and conclusions based on circuit models need to be treated with caution. One attempt to avoid this difficulty involves assuming that the magnetic structure remains fixed and current is transferred between flux tubes in a quadrupolar model \citep{M97,HMH98}.

\begin{figure}
\begin{center}
 \includegraphics[width=0.8\hsize]{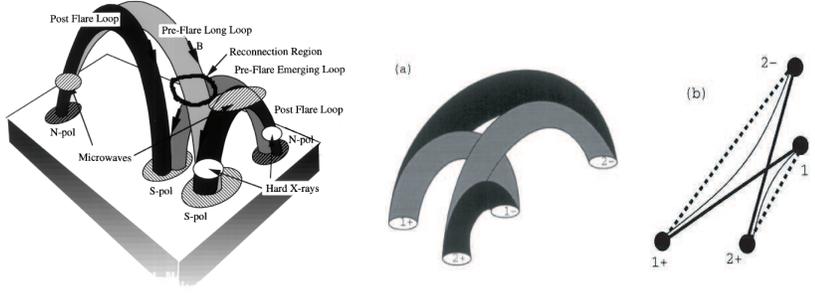}
 \caption{Cartoons illustrating quadrupolar models: left from \citet{Netal97} and middle from \citet{M97}, with the lightly and darkly shaded loops corresponding to pre- and post-flare loops, respectively; the figure on the right shows a projection of the middle cartoon onto the solar surface, with the pre-flare and post-flare loops represented by solid and dashed lines, respectively, and with the faint lines representing the paths of reconnected flux tubes in motion between their pre-flare (solid) and post-flare (dashed) locations.}
 \end{center}
 \label{fig:quadrupolar}
   \end{figure}

\subsection{Quadrupolar models}

In a quadrupolar model, which consists of four footpoints and four loops connecting them, reconnection can allow transfer of magnetic flux and current between the loops, cf.\ Figure~\ref{fig:quadrupolar}, and also the right hand part of Figure~\ref{fig:G48}. Assuming that the $I_i$ do not change during a flare, the only changes are in the geometry, corresponding to changes in the current paths, implying a change in $L_i,M_{ij}\to L'_i,M'_{ij}$. Energy is available to drive the flare if $E'_{\rm mag}<E_{\rm mag}$, with $E'_{\rm mag}$ defined by equation (\ref{Ecm}) with $M_{ij}\to M'_{ij}$.  As in a single-loop model, favorable conditions for the release of magnetic energy involve a reduction in the total current path, such as occurs in  shrinkage of the magnetic/current pattern in a single loop following reconnection. However, a reduction in the net current path is not essential: a change in the orientation of loops without any change in shape can lead to magnetic energy release through a change in mutual inductances \citep{Ketal09}.

A version of the quadrupole model has been used to identify the most favorable  configurations for release of magnetic energy \citep{M97,HMH98,Aetal99}. In this version the geometry is fixed, with the four flux tubes shown in Figure~\ref{fig:quadrupolar} not changing in shape or location during the flare; the only change is in the way the current is distributed between the four flux tubes. Let the currents in the flux loops between footpoints (1+, 1-), (2+,2-), (1+, 2-), (2+,1-) be, respectively, $I_1,I_2,I_3,I_4$ pre-flare and $I'_1,I'_2,I'_3,I'_4$ post-flare. The boundary condition that the current at each of the four footpoints not change requires $I_1+I_3=I'_1+I'_3$ at footpoint 1+, and analogous relations at the other three footpoints. These boundary conditions require that the current, $\Delta I_i=I_i-I'_i$, gained or lost at each of the four footpoints be the same apart from sign: $\Delta I_1=\Delta I_2=-\Delta I_3=-\Delta I_4=\Delta I$ say. The current transfer, $\Delta I$, in this model changes the magnitudes of the currents in each of the four loops, without changing the current at each of the four footpoints. Favorable conditions for maximum magnetic energy release \citep{HMH98,Aetal99} may be summarized qualitatively as configurations that favor current transfer from longer to shorter loops. This version of a quadrupolar model does not include any intrinsic time-dependence: it is based on assumed pre- and post-flare configurations, and does not include the time-dependent processes involved in the transition between these configurations. 

\subsection{Time-dependent inductances}

The discrete model, cf.\ equation (\ref{Ecm}), may be used to discuss some aspects of the time dependence during a flare \citep{Ketal09}. The magnetic energy must decrease, and if the currents are all assumed constant in magnitude during a flare, the time dependence is in the geometric configuration of the current paths, described by time-dependent $M_{ij}$. The magnetic flux associated with the $i$th circuit is $\Psi_i=\sum_jM_{ij}I_j$, and the rate of change of this magnetic flux gives the EMF in the $i$th circuit
\begin{equation}
\Phi_i=-{\dot\Psi}_i=-\sum_j{\dot M}_{ij}I_j=-{\dot L}_iI_i-\sum_{j\ne i}{\dot M}_{ij}I_j,
\label{flux}
\end{equation}
where a dot denotes a time derivative. The power in a flare is then described by 
\begin{equation}
{\dot E}_{\rm mag}=-\sum_i\Phi_iI_i=-{1\over2}\sum_i{\dot L}_iI_i^2-\sum_{j\ne i}{\dot M}_{ij}I_iI_j.
\label{dotEmag}
\end{equation} 


In the single-loop model discussed above, the rate of change of the magnetic energy is $-{1\over2}{\dot L}_iI_i^2$. Noting that ${\dot L}_i$ has the dimensions of a resistance, it might be tempting to identify it as an effective resistance that describes the implicit energy sink. Conversely, given a specific model for the conversion of magnetic energy into other forms, the power $-{1\over2}{\dot L}_iI_i^2$ can be equated to the power appearing in these other forms. One can then identify $-{1\over2}{\dot L}_i$ as the effective resistance for this conversion process.

 As remarked above, the flare-associated current is driven by the EMF, such that the power released is identified as the rate at which this current does work against the EMF, but the path of this current is not known. A constraint of the path of the flare-associated current is implicit in a multiple-loop model with fixed currents and time-varying current paths. As indicated by the faint lines in the right-hand diagram in Figure~\ref{fig:quadrupolar}, the reconfiguration of the current is attributed to magnetic reconfiguration of current-carrying magnetic field lines, which propagate from their initial to their final locations during the flare. There are several qualitative implications that need to be included in any future detailed version of this model. 
\begin{itemize}
\item The flare-associated current involves redirection of existing currents, rather than the generation of an intrinsically new current, implying that its  maximum magnitude is of the same order as the pre-existing current, $I\approx10^{11}\,$A. 
\item The reconfiguration involves motions of magnetic flux tubes, and the speed of such motion is less than the Alfv\'en speed $v_A$, implying that the rate of change of the length of a flux tube, $\ell$, can be written as ${\dot\ell}=\zeta v_A$, with $\zeta\lesssim1$. 
\item The  rate of reduction in the stored magnetic energy is then dissipation-like with the resistance replaced by $\zeta \mu_0v_A$, where $\mu_0v_A$ is sometimes called the Alfv\'en impedance, by analogy with the vacuum impedance $\mu_0c$.
\item Effective dissipation of the flare-associated current requires that it be redirected so that it closes across field lines in the chromosphere, similar to the current-wedge model for the current in a substorm \citep{MRA73}.
\end{itemize}
This final dot point is related to the supply of electrons for acceleration.  As in the current-wedge model, the energy is assumed to be stored in a large volume, released through reconnection in a localized region, with the released energy converted into Alfv\'enic form \citep{Betal99} and transported downward until it is transferred to energetic electrons. The acceleration of the electrons is attributed to a parallel electric field \citep[e.g.,][]{H12,M12a} in an acceleration region that is separate from the current-closure region in the partially ionized plasma below.  The long-standing ``number problem'' \citep{B71,B76} requires that during the flare the corona be continuously resupplied with electrons from the chromosphere replacing the precipitating accelerated electrons. The high density of the precipitating electrons \citep{Ketal11} is consistent with a model in which the source of the accelerated electrons is upflow from the chromosphere. The acceleration is due to the EMF becoming localized along field lines on the upward current paths; the total $\Phi=10^{10}\,$V separates into potential changes of $10^4\,$V along $10^6$ upward current paths \citep{H85,MW13}, and the electrons are accelerated by the resulting $E_\parallel$ just above the chromosphere \citep{H12,M12b,MW13} on each upward current path, analogous to the acceleration of auroral electrons just above the ionosphere. The effective rate of dissipation is then identified in terms of the power going into electrons through the acceleration by $E_\parallel$.

\section{Helicity in flares and CMEs}
\label{sect:helicity}

Magnetic helicity is an important concept in flare physics: it can be estimated quantitatively from observation \citep[e.g.,][]{vDGetal03}, and it is conserved during magnetic reconnection \citep{M69,BF84}. Conservation of magnetic helicity  provides an additional constraint on the currents during magnetic energy release in a flare. It was suggested \citep{M04} that this condition can be imposed in a multiple-current model by relating the self and mutual helicities to the self and mutual inductances in a force-free model, but the assumed relation between helicities and inductances was shown to be incorrect \citep{DPB06}. Corrected forms for the helicity, the helicity budget of the Sun and the helicity in CMEs are discussed in this section.

\begin{figure}
\begin{center}
 \includegraphics[width=0.9\hsize]{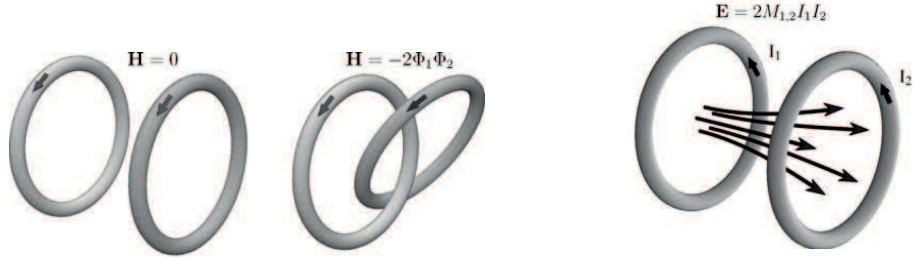}
\caption{The figure on the left shows two  unlinked flux loops, with zero mutual helicity, the middle figure shows two linked flux loops with nonzero mutual helicity; the figure on the right indicates that the mutual inductance of two unlinked flux tubes is nonzero; the arrows on the rings indicate the direction of the current. From \citet{DPB06}.}
\label{fig:helicity1}
\end{center}
  \end{figure}

\subsection{Magnetic helicity for force-free currents}

The magnetic helicity, $H$, may be written as a double integral analogous to the magnetic energy in equation~(\ref{Edef}):
\begin{equation}
H=\int d^3{\bf x}\,{\bf B}({\bf x})\cdot{\bf A}({\bf x}),
\qquad
{\bf A}({\bf x})=\frac{1}{4\pi}\int d^3{\bf x}'\,{\bf B}({\bf x}')\times\frac{({\bf x}-{\bf x}')}
{|{\bf x}-{\bf x}'|^3},
\label{Hdef}
\end{equation}
where the latter relation allows one to express $H$ in terms of ${\bf B}$ \citep{M69,DPB06}:
\begin{equation}
H=\frac{1}{4\pi}\int d^3{\bf x}\int d^3{\bf x}'\,
{\bf B}({\bf x})\times{\bf B}({\bf x}')\cdot\frac{({\bf x}-{\bf x}')}{|{\bf x}-{\bf x}'|^3}.
\label{H2}
\end{equation} 
The force-free assumption implies
\begin{equation}
{\bf B}({\bf x})=\frac{\mu_0{\bf J}({\bf x})}{\alpha({\bf x})},
\label{ffc}
\end{equation}
with $\alpha({\bf x})$ constant along each current line. The helicity for a force-free field follows by substituting equation (\ref{ffc}) into the relation (\ref{H2}).

\citet{DPB06} discussed the evaluation of $H$ for a collection of flux tubes within a finite volume, treating closed and open flux tubes separately. For a collection of flux tubes that close within the volume, equation (\ref{H2}) can be reduced to \citep{BF84}
\begin{equation}
H=\sum_iT_i\Psi_i^2+\sum_{i\ne j}{\cal L}_{ij}\Psi_i\Psi_j,
\label{H3}
\end{equation} 
where $\Psi_i$ is the magnetic flux in the $i$th flux tube, and with $T_i$ and ${\cal L}_{ij}$ identified as the self and mutual helicities. Although equations (\ref{Ecm}) and  (\ref{H3}) have similar forms, the inductances and helicities are not related, even for a force-free magnetic field. A simple counter-example is shown in Figure~\ref{fig:helicity1}.

The self helicity in a single flux loop can be written \citep{BF84} $H=(T+W)\Phi_{\rm mag}^2$, where $T$ is the twist (of magnetic field lines about the axis of flux tube) and $W$ is the writhe  which is associated with the shape of the axis of the flux tube \citep{BP06}. During reconnection, twist can be partly converted into writhe with the total helicity conserved. The conversion of twist into writhe occurs in the kink instability, in which the writhe of the axis of the flux tube grows exponentially, underlying the kink-instability model for the formation of a CME  \citep{S76,TKT04,Fan05}.


\subsection{Helicity budget}

Rising current-carrying flux tubes transport magnetic energy and helicity through the corona into the solar wind.  Over a solar cycle it is estimated  that $H=10^{47}\rm\,Mx^2$ is injected into the corona (the units of $H$ are the same as those of magnetic flux squared, Wb$^2$, and in the solar literature are usually expressed in gaussian units, Mx$^2$, with $1\rm\,Mx=10^{-8}\,Wb$)  and that $10^{45}\rm\,Mx^2$ is transported away by CMEs \citep{BR00}. The injection of helicity is predominantly through the emergence of new current-carrying flux tubes. 
There is an upper bound on the helicity that can accumulate in a flux loop \citep{ZFL06}, leading to the suggestion that the build-up of $H$ triggers a CME when this bound is exceeded. A specific instability is needed to account for the sudden release of a CME, and either the kink instability \citep{S76,TKT04,Fan05} or the torus instability \citep{A78,KT06,Zetal14,Zetal15,Metal17} is favored.

\section{Discussion}
\label{sect:discussion}

The role of currents in the solar flares and CMEs that follows from the foregoing discussion may be summarized as follows:
\begin{itemize}
\item The ``free'' magnetic energy available for release in a solar flare is stored in currents flowing in coronal magnetic loops.
\item The energetically important currents  flow from one footpoint of the loop to the other, and  close  deep inside the Sun.
\item Release of stored magnetic energy is due to reconfiguration of the current system, resulting in a net reduction in the effective current path (a magnetic ``shrinkage''), rather than a net reduction in the current (a ``current interruption'').
\item Currents flow predominantly along field lines, and a reconfiguration of the current system is associated with a reconfiguration of the magnetic field.
\item A flare is magnetically driven: the EMF due to the changing magnetic flux drives a flare-associated current, and the power released is equal to the rate work is done by the EMF against this current; the driver may also be interpreted as the Lorentz force doing work against a fluid flow.
\item The net effect of the flare-associated current over the duration of the flare is to change the pre-flare current configuration into the post-flare current configuration.
\item  Cross-field current flow is ineffective in closing the flare-associated current in the corona; the current is redirected along field lines to the chromosphere and back, closing across chromospheric field lines due to the Pedersen conductivity.
\item The EMF localizes along field lines on a large number of up-current paths above the chromosphere, resulting in the ``bulk energization'' of 10--20$\,$keV electrons that produce the various signatures of the impulsive phase.
\item  The formation of a CME can be triggered by the kink instability, with the subsequent acceleration of the CME driven by the repulsive current-current force in the current loop, e.g., as in the torus instability.
\item Magnetic helicity can  build up in a prominence until a threshold is reached, resulting in the instability that leads to a CME that carries away the excess helicity.
\end{itemize}

Such a current-based model for solar flares can be simplified and reduced to a small number of parameters, allowing scaling of the model. Such scaling is relevant not only for solar flares, which occur with a wide range of energy output, but also for flare stars, superflares on solar-like stars and other examples of magnetic explosions.  Ignoring numerical factors of order unity, the stored magnetic energy can be approximated by $\mu_0\ell I^2$, where $\ell$ is a scale length and $I$ is the current flowing through the corona.  The energy released in a magnetic shrinkage at constant $I$ is then proportional to $\mu_0\Delta\ell I^2$, where $-\Delta\ell$ is the change in $\ell$ that results from the shrinkage.  Assuming $\Delta\ell\propto\ell$, the characteristic time scale for the energy release is $T=\ell/v_A$. An additional assumption is required to determine how the current scales. The current must be confined by the potential magnetic field: for example, in a cylindrical model for a flux tube of radius $R$ this implies that the maximum current scales as $I=4\pi RB/\mu_0$. Assuming that the current is close to this maximum value is the additional assumption needed. The available magnetic  free energy is $\mu_0\Delta\ell I^2=\mu_0\Delta\ell(4\pi RB/\mu_0)^2$ for this maximal current, and this scales $\propto(B^2/2\mu_0)R^2\ell$, as expected in a model that emphasizes $B$ rather than $I$.  If one makes the further scaling $R\propto\ell$, this simple version of the model involves only the parameters $\ell,I,v_A$. The wide range of energy, $\mu_0\ell I^2$, is presumably due to a wide range of $I$ in different flares. However, it is in connection with speculations about the largest flares that such scaling is of most interest.

Consider a superflare on a solar-type star involving an energy release two to three orders of magnitude larger than in a large solar flare. Suppose that the $B$ in emerging flux is fixed, as is approximately the case in all emerging flux on the Sun ($B\approx0.1\,$T). Assuming maximal current, the energy release scales as $B^2R^2\ell\propto B^2\ell^3$, which is the same as the scaling inferred by \citet{Setal13} based on scaling of the magnetic energy. For example, a superflare requires a starspot an order of magnitude larger than a typical sunspot. Assuming $v_A$ is the same as in the solar corona, the duration of the energy release scales as $\ell$, so that the duration of such a superflare would be an order of magnitude longer than a solar flare. 

Another example of a magnetic explosion is an outburst on a magnetar  \citep{DT92,WT06}; existing models for such outbursts are based on scaled models for solar flare \citep{L03,L06,Petal12,Eetal16}.  Scaling the model envisaged here, assuming a surface magnetic field of order $10^{10}\,$T, a scale length $\ell=10^3\,$m of order one tenth the stellar radius, the scaling factor for the maximum current of order $10^7$ implying a scaling factor for the energy is of order $10^{14}$ compared with a solar flare, whereas a more realistic value is of order $10^{10}$. This suggests that the currents involved in a magnetar outburst are only a small fraction ($\approx1$\%) of the maximum current that can be confined by the magnetic field. However, the application of any flare model to magnetar outbursts needs further critical discussion, and the scaling suggested here is only indicative.

\acknowledgments
I thank Mike Wheatland and Alpha Mastrano for helpful comments on the manuscript.

This paper is theoretical and contains neither new data nor new models.

\bibliographystyle{agufull08}

\begin{thebibliography}{105}
\providecommand{\natexlab}[1]{#1}
\expandafter\ifx\csname urlstyle\endcsname\relax
  \providecommand{\doi}[1]{doi:\discretionary{}{}{}#1}\else
  \providecommand{\doi}{doi:\discretionary{}{}{}\begingroup
  \urlstyle{rm}\Url}\fi

\bibitem[{\textit{{Alfv{\'e}n} and {Carlqvist}}(1967)}]{AC67}
{Alfv{\'e}n}, H., and P.~{Carlqvist} (1967), {Currents in the Solar Atmosphere
  and a Theory of Solar Flares}, \textit{\solphys}, \textit{1}, 220--228,
  \doi{10.1007/BF00150857}.

\bibitem[{\textit{{Anzer}}(1978)}]{A78}
{Anzer}, U. (1978), {Can coronal loop transients be driven magnetically},
  \textit{\solphys}, \textit{57}, 111--118, \doi{10.1007/BF00152048}.

\bibitem[{\textit{{Aschwanden}}(2004)}]{A04}
{Aschwanden}, M.~J. (2004), \textit{{Physics of the Solar Corona. An
  Introduction}}, Praxis Publishing Ltd.

\bibitem[{\textit{{Aschwanden} et~al.}(1999)\textit{{Aschwanden}, {Kosugi},
  {Hanaoka}, {Nishio}, and {Melrose}}}]{Aetal99}
{Aschwanden}, M.~J., T.~{Kosugi}, Y.~{Hanaoka}, M.~{Nishio}, and D.~B.
  {Melrose} (1999), {Quadrupolar Magnetic Reconnection in Solar Flares. I.
  Three-dimensional Geometry Inferred from Yohkoh Observations}, \textit{\apj},
  \textit{526}, 1026--1045, \doi{10.1086/308025}.

\bibitem[{\textit{{Atkinson}}(1970)}]{A70}
{Atkinson}, G. (1970), {Auroral arcs: Result of the interaction of a dynamic
  magnetosphere with the ionosphere}, \textit{\jgr}, \textit{75}, 4746,
  \doi{10.1029/JA075i025p04746}.

\bibitem[{\textit{{Aulanier} et~al.}(2005)\textit{{Aulanier}, {D{\'e}moulin},
  and {Grappin}}}]{ADG05}
{Aulanier}, G., P.~{D{\'e}moulin}, and R.~{Grappin} (2005), {Equilibrium and
  observational properties of line-tied twisted flux tubes}, \textit{\aap},
  \textit{430}, 1067--1087, \doi{10.1051/0004-6361:20041519}.
  

\bibitem[{\textit{{Barnes} et~al.}(2016)\textit{{Barnes}, {Leka}, {Schrijver},
  {Colak}, {Qahwaji}, {Ashamari}, {Yuan}, {Zhang}, {McAteer}, {Bloomfield},
  {Higgins}, {Gallagher}, {Falconer}, {Georgoulis}, {Wheatland}, {Balch},
  {Dunn}, and {Wagner}}}]{Betal16}
{Barnes}, G., K.~D. {Leka}, C.~J. {Schrijver}, T.~{Colak}, R.~{Qahwaji}, O.~W.
  {Ashamari}, Y.~{Yuan}, J.~{Zhang}, R.~T.~J. {McAteer}, D.~S. {Bloomfield},
  P.~A. {Higgins}, P.~T. {Gallagher}, D.~A. {Falconer}, M.~K. {Georgoulis},
  M.~S. {Wheatland}, C.~{Balch}, T.~{Dunn}, and E.~L. {Wagner} (2016), {A
  Comparison of Flare Forecasting Methods. I. Results from the ``All-Clear''
  Workshop}, \textit{\apj}, \textit{829}, 89, \doi{10.3847/0004-637X/829/2/89}.

\bibitem[{\textit{{Bateman}}(1978)}]{B78}
{Bateman}, G. (1978), \textit{{MHD instabilities}}, MIT Press.

\bibitem[{\textit{{Berger} and {Field}}(1984)}]{BF84}
{Berger}, M.~A., and G.~B. {Field} (1984), {The topological properties of
  magnetic helicity}, \textit{Journal of Fluid Mechanics}, \textit{147},
  133--148, \doi{10.1017/S0022112084002019}.

\bibitem[{\textit{{Berger} and {Prior}}(2006)}]{BP06}
{Berger}, M.~A., and C.~{Prior} (2006), {The writhe of open and closed curves},
  \textit{Journal of Physics A Mathematical General}, \textit{39}, 8321--8348,
  \doi{10.1088/0305-4470/39/26/005}.
 

\bibitem[{\textit{{Berger} and {Ruzmaikin}}(2000)}]{BR00}
{Berger}, M.~A., and A.~{Ruzmaikin} (2000), {Rate of helicity production by
  solar rotation}, \textit{\jgr}, \textit{105}, 10,481--10,490,
  \doi{10.1029/1999JA900392}.

\bibitem[{\textit{{Birn} and {Hesse}}(1996)}]{BH96}
{Birn}, J., and M.~{Hesse} (1996), {Details of current disruption and diversion
  in simulations of magnetotail dynamics}, \textit{\jgr}, \textit{101},
  15,345--15,358, \doi{10.1029/96JA00887}.

\bibitem[{\textit{{Birn} et~al.}(1999)\textit{{Birn}, {Hesse}, {Haerendel},
  {Baumjohann}, and {Shiokawa}}}]{Betal99}
{Birn}, J., M.~{Hesse}, G.~{Haerendel}, W.~{Baumjohann}, and K.~{Shiokawa}
  (1999), {Flow braking and the substorm current wedge}, \textit{\jgr},
  \textit{104}, 19,895--19,904, \doi{10.1029/1999JA900173}.

\bibitem[{\textit{{Brown}}(1971)}]{B71}
{Brown}, J.~C. (1971), {The Deduction of Energy Spectra of Non-Thermal
  Electrons in Flares from the Observed Dynamic Spectra of Hard X-Ray Bursts},
  \textit{\solphys}, \textit{18}, 489--502, \doi{10.1007/BF00149070}.

\bibitem[{\textit{{Brown}}(1976)}]{B76}
{Brown}, J.~C. (1976), {The interpretation of hard and soft X-rays from solar
  flares}, \textit{Philosophical Transactions of the Royal Society of London
  Series A}, \textit{281}, 473--490, \doi{10.1098/rsta.1976.0044}.

\bibitem[{\textit{{Brown} and {Bingham}}(1984)}]{BB84}
{Brown}, J.~C., and R.~{Bingham} (1984), {Electrodynamics effects in
  beam/return current systems and their implications for solar impulsive
  bursts}, \textit{\aap}, \textit{131}, L11--L14.

\bibitem[{\textit{{Brown} and {Melrose}}(1977)}]{BM77}
{Brown}, J.~C., and D.~B. {Melrose} (1977), {Collective plasma effects and the
  electron number problem in solar hard X-ray bursts}, \textit{\solphys},
  \textit{52}, 117--131, \doi{10.1007/BF00935795}.

\bibitem[{\textit{{Carmichael}}(1964)}]{C64}
{Carmichael}, H. (1964), {A Process for Flares}, \textit{NASA Special
  Publication}, \textit{50}, 451.

\bibitem[{\textit{{Compagnino} et~al.}(2017)\textit{{Compagnino}, {Romano}, and
  {Zuccarello}}}]{CRZ17}
{Compagnino}, A., P.~{Romano}, and F.~{Zuccarello} (2017), {A Statistical Study
  of CME Properties and of the Correlation Between Flares and CMEs over Solar
  Cycles 23 and 24}, \textit{\solphys}, \textit{292}, 5,
  \doi{10.1007/s11207-016-1029-4}.

\bibitem[{\textit{{Dalmasse} et~al.}(2015)\textit{{Dalmasse}, {Aulanier},
  {D{\'e}moulin}, {Kliem}, {T{\"o}r{\"o}k}, and {Pariat}}}]{Detal15}
{Dalmasse}, K., G.~{Aulanier}, P.~{D{\'e}moulin}, B.~{Kliem},
  T.~{T{\"o}r{\"o}k}, and E.~{Pariat} (2015), {The Origin of Net Electric
  Currents in Solar Active Regions}, \textit{\apj}, \textit{810}, 17,
  \doi{10.1088/0004-637X/810/1/17}.
 

\bibitem[{\textit{{D{\'e}moulin} and {Aulanier}}(2010)}]{DA10}
{D{\'e}moulin}, P., and G.~{Aulanier} (2010), {Criteria for Flux Rope Eruption:
  Non-equilibrium Versus Torus Instability}, \textit{\apj}, \textit{718},
  1388--1399, \doi{10.1088/0004-637X/718/2/1388}.

\bibitem[{\textit{{D{\'e}moulin} et~al.}(2006)\textit{{D{\'e}moulin}, {Pariat},
  and {Berger}}}]{DPB06}
{D{\'e}moulin}, P., E.~{Pariat}, and M.~A. {Berger} (2006), {Basic Properties
  of Mutual Magnetic Helicity}, \textit{\solphys}, \textit{233}, 3--27,
  \doi{10.1007/s11207-006-0010-z}.

\bibitem[{\textit{{Duncan} and {Thompson}}(1992)}]{DT92}
{Duncan}, R.~C., and C.~{Thompson} (1992), {Formation of very strongly
  magnetized neutron stars - Implications for gamma-ray bursts},
  \textit{\apjl}, \textit{392}, L9--L13, \doi{10.1086/186413}.

\bibitem[{\textit{{Dungey}}(1953)}]{D53}
{Dungey}, J.~W. (1953), {The motion of magnetic fields}, \textit{\mnras},
  \textit{113}, 679, \doi{10.1093/mnras/113.6.679}.

\bibitem[{\textit{{Dungey}}(1958)}]{D58}
{Dungey}, J.~W. (1958), {The Neutral Point Discharge Theory of Solar Flares. a
  Reply to Cowling's Criticism}, in \textit{Electromagnetic Phenomena in
  Cosmical Physics}, \textit{IAU Symposium}, vol.~6, edited by B.~{Lehnert}, p.
  135.

\bibitem[{\textit{{Elenbaas} et~al.}(2016)\textit{{Elenbaas}, {Watts},
  {Turolla}, and {Heyl}}}]{Eetal16}
{Elenbaas}, C., A.~L. {Watts}, R.~{Turolla}, and J.~S. {Heyl} (2016), {The
  impulsive phase of magnetar giant flares: assessing linear tearing as the
  trigger mechanism}, \textit{\mnras}, \textit{456}, 3282--3295,
  \doi{10.1093/mnras/stv2860}.

\bibitem[{\textit{{Emslie} and {Henoux}}(1995)}]{EH95}
{Emslie}, A.~G., and J.-C. {Henoux} (1995), {The Electrical Current Structure
  Associated with Solar Flare Electrons Accelerated by Large-Scale Electric
  Fields}, \textit{\apj}, \textit{446}, 371, \doi{10.1086/175796}.

\bibitem[{\textit{{Ergun} et~al.}(2002)\textit{{Ergun}, {Andersson}, {Main},
  {Su}, {Newman}, {Goldman}, {Carlson}, {McFadden}, and {Mozer}}}]{Eetal02}
{Ergun}, R.~E., L.~{Andersson}, D.~{Main}, Y.-J. {Su}, D.~L. {Newman}, M.~V.
  {Goldman}, C.~W. {Carlson}, J.~P. {McFadden}, and F.~S. {Mozer} (2002),
  {Parallel electric fields in the upward current region of the aurora:
  Numerical solutions}, \textit{\pop}, \textit{9}, 3695--3704,
  \doi{10.1063/1.1499121}.

\bibitem[{\textit{{Fan}}(2005)}]{Fan05}
{Fan}, Y. (2005), {Coronal Mass Ejections as Loss of Confinement of Kinked
  Magnetic Flux Ropes}, \textit{\apj}, \textit{630}, 543--551,
  \doi{10.1086/431733}.

\bibitem[{\textit{{Fletcher} and {Hudson}}(2008)}]{FH08}
{Fletcher}, L., and H.~S. {Hudson} (2008), {Impulsive Phase Flare Energy
  Transport by Large-Scale Alfv{\'e}n Waves and the Electron Acceleration
  Problem}, \textit{\apj}, \textit{675}, 1645-1655, \doi{10.1086/527044}.
  
  \bibitem[{\textit{{Forbes} and {Acton}}(1996)}]{FA96}
{Forbes}, T.~G., and L.~W. {Acton} (1996), {Reconnection and Field Line
  Shrinkage in Solar Flares}, \textit{\apj}, \textit{459}, 330,
  \doi{10.1086/176896}.

\bibitem[{\textit{{Forbes} and {Malherbe}}(1986)}]{FM86}
{Forbes}, T.~G., and J.~M. {Malherbe} (1986), {A shock condensation mechanism
  for loop prominences}, \textit{\apjl}, \textit{302}, L67--L70,
  \doi{10.1086/184639}.

\bibitem[{\textit{{Georgoulis} et~al.}(2012)\textit{{Georgoulis}, {Titov}, and
  {Miki{\'c}}}}]{Getal12}
{Georgoulis}, M.~K., V.~S. {Titov}, and Z.~{Miki{\'c}} (2012), {Non-neutralized
  Electric Current Patterns in Solar Active Regions: Origin of the
  Shear-generating Lorentz Force}, \textit{\apj}, \textit{761}, 61,
  \doi{10.1088/0004-637X/761/1/61}.

\bibitem[{\textit{{Giovanelli}}(1946)}]{G46}
{Giovanelli}, R.~G. (1946), {A Theory of Chromospheric Flares}, \textit{\nat},
  \textit{158}, 81--82, \doi{10.1038/158081a0}.

\bibitem[{\textit{{Giovanelli}}(1947)}]{G47}
{Giovanelli}, R.~G. (1947), {Magnetic and Electric Phenomena in the Sun's
  Atmosphere associated with Sunspots}, \textit{\mnras}, \textit{107}, 338,
  \doi{10.1093/mnras/107.4.338}.

\bibitem[{\textit{{Giovanelli}}(1948)}]{G48}
{Giovanelli}, R.~G. (1948), {Chromospheric Flares}, \textit{\mnras},
  \textit{108}, 163, \doi{10.1093/mnras/108.2.163}.

\bibitem[{\textit{{Gold} and {Hoyle}}(1960)}]{GH60}
{Gold}, T., and F.~{Hoyle} (1960), {On the origin of solar flares},
  \textit{\mnras}, \textit{120}, 89, \doi{10.1093/mnras/120.2.89}.

\bibitem[{\textit{{Greene}}(1988)}]{G88}
{Greene}, J.~M. (1988), {Geometrical properties of three-dimensional
  reconnecting magnetic fields with nulls}, \textit{\jgr}, \textit{93},
  8583--8590, \doi{10.1029/JA093iA08p08583}.

\bibitem[{\textit{{Haerendel}}(2012)}]{H12}
{Haerendel}, G. (2012), {Solar Auroras}, \textit{\apj}, \textit{749}, 166,
  \doi{10.1088/0004-637X/749/2/166}.

\bibitem[{\textit{{Hardy} et~al.}(1998)\textit{{Hardy}, {Melrose}, and
  {Hudson}}}]{HMH98}
{Hardy}, S.~J., D.~B. {Melrose}, and H.~S. {Hudson} (1998), {Observational
  Tests of a Double Loop Model for Solar Flares}, \textit{\pasa}, \textit{15},
  318--324, \doi{10.1071/AS98318}.

\bibitem[{\textit{{Hirayama}}(1974)}]{H74}
{Hirayama}, T. (1974), {Theoretical Model of Flares and Prominences. I:
  Evaporating Flare Model}, \textit{\solphys}, \textit{34}, 323--338,
  \doi{10.1007/BF00153671}.

\bibitem[{\textit{{Holman}}(1985)}]{H85}
{Holman}, G.~D. (1985), {Acceleration of runaway electrons and Joule heating in
  solar flares}, \textit{\apj}, \textit{293}, 584--594, \doi{10.1086/163263}.

\bibitem[{\textit{{Hudson}}(2016)}]{H16}
{Hudson}, H.~S. (2016), {Chasing White-Light Flares}, \textit{\solphys},
  \textit{291}, 1273--1322, \doi{10.1007/s11207-016-0904-3}.

\bibitem[{\textit{{Jackson}}(1975)}]{J75}
{Jackson}, J.~D. (1975), \textit{{Classical electrodynamics}}, Wiley: New York.

\bibitem[{\textit{{Janvier} et~al.}(2014)\textit{{Janvier}, {Aulanier},
  {Bommier}, {Schmieder}, {D{\'e}moulin}, and {Pariat}}}]{Jetal14}
{Janvier}, M., G.~{Aulanier}, V.~{Bommier}, B.~{Schmieder}, P.~{D{\'e}moulin},
  and E.~{Pariat} (2014), {Electric Currents in Flare Ribbons: Observations and
  Three-dimensional Standard Model}, \textit{\apj}, \textit{788}, 60,
  \doi{10.1088/0004-637X/788/1/60}.
  
  \bibitem[{\textit{{Janvier} et~al.}(2015)\textit{{Janvier}, {Aulanier}, and
  {D{\'e}moulin}}}]{JAD15}
{Janvier}, M., G.~{Aulanier}, and P.~{D{\'e}moulin} (2015), {From Coronal
  Observations to MHD Simulations, the Building Blocks for 3D Models of Solar
  Flares (Invited Review)}, \textit{\solphys}, \textit{290}, 3425--3456,
  \doi{10.1007/s11207-015-0710-3}.
  
  \bibitem[{\textit{{Janvier} et~al.}(2014)\textit{{Janvier}, {Aulanier},
  {Bommier}, {Schmieder}, {D{\'e}moulin}, and {Pariat}}}]{Jetal14}
{Janvier}, M., G.~{Aulanier}, V.~{Bommier}, B.~{Schmieder}, P.~{D{\'e}moulin},
  and E.~{Pariat} (2014), {Electric Currents in Flare Ribbons: Observations and
  Three-dimensional Standard Model}, \textit{\apj}, \textit{788}, 60,
  \doi{10.1088/0004-637X/788/1/60}.


\bibitem[{\textit{{Khodachenko} et~al.}(2009)\textit{{Khodachenko}, {Zaitsev},
  {Kislyakov}, and {Stepanov}}}]{Ketal09}
{Khodachenko}, M.~L., V.~V. {Zaitsev}, A.~G. {Kislyakov}, and A.~V. {Stepanov}
  (2009), {Equivalent Electric Circuit Models of Coronal Magnetic Loops and
  Related Oscillatory Phenomena on the Sun}, \textit{\ssr}, \textit{149},
  83--117, \doi{10.1007/s11214-009-9538-1}.

\bibitem[{\textit{{Kliem} and {T{\"o}r{\"o}k}}(2006)}]{KT06}
{Kliem}, B., and T.~{T{\"o}r{\"o}k} (2006), {Torus Instability}, \textit{\prl},
  \textit{96}(25), 255002, \doi{10.1103/PhysRevLett.96.255002}.

\bibitem[{\textit{{Knudsen}}(1996)}]{K96}
{Knudsen}, D.~J. (1996), {Spatial modulation of electron energy and density by
  nonlinear stationary inertial Alfv{\'e}n waves}, \textit{\jgr}, \textit{101},
  10,761--10,772, \doi{10.1029/96JA00429}.

\bibitem[{\textit{{Kopp} and {Pneuman}}(1976)}]{KP76}
{Kopp}, R.~A., and G.~W. {Pneuman} (1976), {Magnetic reconnection in the corona
  and the loop prominence phenomenon}, \textit{\solphys}, \textit{50}, 85--98,
  \doi{10.1007/BF00206193}.

\bibitem[{\textit{{Krucker} et~al.}(2011)\textit{{Krucker}, {Hudson},
  {Jeffrey}, {Battaglia}, {Kontar}, {Benz}, {Csillaghy}, and {Lin}}}]{Ketal11}
{Krucker}, S., H.~S. {Hudson}, N.~L.~S. {Jeffrey}, M.~{Battaglia}, E.~P.
  {Kontar}, A.~O. {Benz}, A.~{Csillaghy}, and R.~P. {Lin} (2011),
  {High-resolution Imaging of Solar Flare Ribbons and Its Implication on the
  Thick-target Beam Model}, \textit{\apj}, \textit{739}, 96,
  \doi{10.1088/0004-637X/739/2/96}.

\bibitem[{\textit{{Kuperus} and {Raadu}}(1974)}]{KR74}
{Kuperus}, M., and M.~A. {Raadu} (1974), {The Support of Prominences Formed in
  Neutral Sheets}, \textit{\aap}, \textit{31}, 189.

\bibitem[{\textit{{Lau} and {Finn}}(1991)}]{LF91}
{Lau}, Y.-T., and J.~M. {Finn} (1991), {Three-dimensional kinematic
  reconnection of plasmoids}, \textit{\apj}, \textit{366}, 577--591,
  \doi{10.1086/169593}.

\bibitem[{\textit{{Leka} et~al.}(1996)\textit{{Leka}, {Canfield}, {McClymont},
  and {van Driel-Gesztelyi}}}]{Letal96}
{Leka}, K.~D., R.~C. {Canfield}, A.~N. {McClymont}, and L.~{van
  Driel-Gesztelyi} (1996), {Evidence for Current-carrying Emerging Flux},
  \textit{\apj}, \textit{462}, 547, \doi{10.1086/177171}.

\bibitem[{\textit{{Lin} and {Forbes}}(2000)}]{LF00}
{Lin}, J., and T.~G. {Forbes} (2000), {Effects of reconnection on the coronal
  mass ejection process}, \textit{\jgr}, \textit{105}, 2375--2392,
  \doi{10.1029/1999JA900477}.
 

\bibitem[{\textit{{Longcope}}(2005)}]{L05}
{Longcope}, D.~W. (2005), {Topological Methods for the Analysis of Solar
  Magnetic Fields}, \textit{Living Reviews in Solar Physics}, \textit{2},
  \doi{10.12942/lrsp-2005-7}.

\bibitem[{\textit{{Lyutikov}}(2003)}]{L03}
{Lyutikov}, M. (2003), {Explosive reconnection in magnetars}, \textit{\mnras},
  \textit{346}, 540--554, \doi{10.1046/j.1365-2966.2003.07110.x}.

\bibitem[{\textit{{Lyutikov}}(2006)}]{L06}
{Lyutikov}, M. (2006), {Magnetar giant flares and afterglows as relativistic
  magnetized explosions}, \textit{\mnras}, \textit{367}, 1594--1602,
  \doi{10.1111/j.1365-2966.2006.10069.x}.

\bibitem[{\textit{{Mallinckrodt} and {Carlson}}(1978)}]{MC78}
{Mallinckrodt}, A.~J., and C.~W. {Carlson} (1978), {Relations between
  transverse electric fields and field-aligned currents}, \textit{\jgr},
  \textit{83}, 1426--1432, \doi{10.1029/JA083iA04p01426}.

\bibitem[{\textit{{Maltsev} et~al.}(1977)\textit{{Maltsev}, {Liatskii}, and
  {Liatskaia}}}]{Metal77}
{Maltsev}, I.~P., V.~B. {Liatskii}, and A.~M. {Liatskaia} (1977), {Currents
  over the auroral arc}, \textit{\planss}, \textit{25}, 53--57,
  \doi{10.1016/0032-0633(77)90117-9}.

\bibitem[{\textit{{Martens} and {Kuin}}(1989)}]{MK89}
{Martens}, P.~C.~H., and N.~P.~M. {Kuin} (1989), {A circuit model for filament
  eruptions and two-ribbon flares}, \textit{\solphys}, \textit{122}, 263--302,
  \doi{10.1007/BF00912996}.
  

\bibitem[{\textit{{McPherron} et~al.}(1973)\textit{{McPherron}, {Russell}, and
  {Aubry}}}]{MRA73}
{McPherron}, R.~L., C.~T. {Russell}, and M.~P. {Aubry} (1973), {Satellite
  studies of magnetospheric substorms on August 15, 1968: 9. Phenomenological
  model for substorms}, \textit{\jgr}, \textit{78}, 3131,
  \doi{10.1029/JA078i016p03131}.

\bibitem[{\textit{{Melrose}}(2004)}]{M04}
{Melrose}, D. (2004), {Conservation of both current and helicity in a
  quadrupolar model for solar flares}, \textit{\solphys}, \textit{221},
  121--133, \doi{10.1023/B:SOLA.0000033358.64885.3a}.

\bibitem[{\textit{{Melrose}}(1992)}]{M92}
{Melrose}, D.~B. (1992), {Energy propagation into a flare kernel during a solar
  fire}, \textit{\apj}, \textit{387}, 403--413, \doi{10.1086/171092}.

\bibitem[{\textit{{Melrose}}(1995)}]{M95}
{Melrose}, D.~B. (1995), {Current Paths in the Corona and Energy Release in
  Solar Flares}, \textit{\apj}, \textit{451}, 391, \doi{10.1086/176228}.

\bibitem[{\textit{{Melrose}}(1996)}]{M96}
{Melrose}, D.~B. (1996), {Reply to Comments by E. N. Parker}, \textit{\apj},
  \textit{471}, 497, \doi{10.1086/177985}.

\bibitem[{\textit{{Melrose}}(1997)}]{M97}
{Melrose}, D.~B. (1997), {A Solar Flare Model Based on Magnetic Reconnection
  between Current-carrying Loops}, \textit{\apj}, \textit{486}, 521--533,
  \doi{10.1086/304521}.

\bibitem[{\textit{{Melrose}}(2012{\natexlab{a}})}]{M12a}
{Melrose}, D.~B. (2012{\natexlab{a}}), {Magnetic Explosions: Role of the
  Inductive Electric Field}, \textit{\apj}, \textit{749}, 59,
  \doi{10.1088/0004-637X/749/1/59}.

\bibitem[{\textit{{Melrose}}(2012{\natexlab{b}})}]{M12b}
{Melrose}, D.~B. (2012{\natexlab{b}}), {Generic Model for Magnetic Explosions
  Applied to Solar Flares}, \textit{\apj}, \textit{749}, 58,
  \doi{10.1088/0004-637X/749/1/58}.

\bibitem[{\textit{{Melrose} and {Wheatland}}(2013)}]{MW13}
{Melrose}, D.~B., and M.~S. {Wheatland} (2013), {Transfer of Energy, Potential,
  and Current by Alfv{\'e}n Waves in Solar Flares}, \textit{\solphys},
  \textit{288}, 223--240, \doi{10.1007/s11207-013-0305-9}.

\bibitem[{\textit{{Melrose} and {Wheatland}}(2014)}]{MW14}
{Melrose}, D.~B., and M.~S. {Wheatland} (2014), {Bulk Energization of Electrons
  in Solar Flares by Alfv{\'e}n Waves}, \textit{\solphys}, \textit{289},
  881--897, \doi{10.1007/s11207-013-0376-7}.

\bibitem[{\textit{{Melrose} et~al.}(1994)\textit{{Melrose}, {Nicholls}, and
  {Broderick}}}]{MNB94}
{Melrose}, D.~B., J.~{Nicholls}, and N.~G. {Broderick} (1994), {Surface
  currents on models of force-free solar magnetic flux tubes}, \textit{Journal
  of Plasma Physics}, \textit{51}, 163--176, \doi{10.1017/S0022377800017451}.

\bibitem[{\textit{{Moffatt}}(1969)}]{M69}
{Moffatt}, H.~K. (1969), {The degree of knottedness of tangled vortex lines},
  \textit{Journal of Fluid Mechanics}, \textit{35}, 117--129,
  \doi{10.1017/S0022112069000991}.

\bibitem[{\textit{{Myers} et~al.}(2017)\textit{{Myers}, {Yamada}, {Ji}, {Yoo},
  {Jara-Almonte}, and {Fox}}}]{Metal17}
{Myers}, C.~E., M.~{Yamada}, H.~{Ji}, J.~{Yoo}, J.~{Jara-Almonte}, and W.~{Fox}
  (2017), {Quasi-static and dynamic magnetic tension forces in arched,
  line-tied magnetic flux ropes}, \textit{Plasma Physics and Controlled
  Fusion}, \textit{59}(1), 014048, \doi{10.1088/0741-3335/59/1/014048}.

\bibitem[{\textit{{Nishio} et~al.}(1997)\textit{{Nishio}, {Yaji}, {Kosugi},
  {Nakajima}, and {Sakurai}}}]{Netal97}
{Nishio}, M., K.~{Yaji}, T.~{Kosugi}, H.~{Nakajima}, and T.~{Sakurai} (1997),
  {Magnetic Field Configuration in Impulsive Solar Flares Inferred from
  Coaligned Microwave/X-Ray Images}, \textit{\apj}, \textit{489}, 976--991,
  \doi{10.1086/304793}.

\bibitem[{\textit{{Parfrey} et~al.}(2012)\textit{{Parfrey}, {Beloborodov}, and
  {Hui}}}]{Petal12}
{Parfrey}, K., A.~M. {Beloborodov}, and L.~{Hui} (2012), {Twisting,
  Reconnecting Magnetospheres and Magnetar Spindown}, \textit{\apjl},
  \textit{754}, L12, \doi{10.1088/2041-8205/754/1/L12}.

\bibitem[{\textit{{Parker}}(1957)}]{P57}
{Parker}, E.~N. (1957), {Sweet's Mechanism for Merging Magnetic Fields in
  Conducting Fluids}, \textit{\jgr}, \textit{62}, 509--520,
  \doi{10.1029/JZ062i004p00509}.

\bibitem[{\textit{{Parker}}(1996)}]{P96}
{Parker}, E.~N. (1996), {Comment on ``Current Paths in the Corona and Energy
  Release in Solar Flares''}, \textit{\apj}, \textit{471}, 489,
  \doi{10.1086/177984}.

\bibitem[{\textit{{Parnell} et~al.}(2015)\textit{{Parnell}, {Stevenson},
  {Threlfall}, and {Edwards}}}]{Petal15}
{Parnell}, C.~E., J.~E.~H. {Stevenson}, J.~{Threlfall}, and S.~J. {Edwards}
  (2015), {Is magnetic topology important for heating the solar atmosphere?},
  \textit{Philosophical Transactions of the Royal Society of London Series A},
  \textit{373}, 20140,264--20140,264, \doi{10.1098/rsta.2014.0264}.

\bibitem[{\textit{{Petschek}}(1964)}]{P64}
{Petschek}, H.~E. (1964), {Magnetic Field Annihilation}, \textit{NASA Special
  Publication}, \textit{50}, 425.

\bibitem[{\textit{{Pontin}}(2011)}]{P11}
{Pontin}, D.~I. (2011), {Three-dimensional magnetic reconnection regimes: A
  review}, \textit{Advances in Space Research}, \textit{47}, 1508--1522,
  \doi{10.1016/j.asr.2010.12.022}.
  
  \bibitem[{\textit{{Priest} and {Forbes}}(2002)}]{PF02}
{Priest}, E.~R., and T.~G. {Forbes} (2002), {The magnetic nature of solar
  flares}, \textit{\aapr}, \textit{10}, 313--377, \doi{10.1007/s001590100013}.


\bibitem[{\textit{{Sakurai}}(1976)}]{S76}
{Sakurai}, T. (1976), {Magnetohydrodynamic interpretation of the motion of
  prominences}, \textit{\pasj}, \textit{28}, 177--198.

\bibitem[{\textit{{Sato} and {Holzer}}(1973)}]{SH73}
{Sato}, T., and T.~E. {Holzer} (1973), {Quiet auroral arcs and electrodynamic
  coupling between the ionosphere and the magnetosphere, 1}, \textit{\jgr},
  \textit{78}, 7314, \doi{10.1029/JA078i031p07314}.

\bibitem[{\textit{{Shafranov}}(1966)}]{Sh66}
{Shafranov}, V.~D. (1966), {Plasma Equilibrium in a Magnetic Field},
  \textit{Reviews of Plasma Physics}, \textit{2}, 103.
  
  \bibitem[{\textit{{Shibata}}(2005)}]{Shibata05}
{Shibata}, K. (2005), {Theories of Eruptive Flares}, in \textit{Coronal and
  Stellar Mass Ejections}, \textit{IAU Symposium}, vol. 226, edited by
  K.~{Dere}, J.~{Wang}, and Y.~{Yan}, pp. 241--249,
  \doi{10.1017/S1743921305000645}.

\bibitem[{\textit{Shibata et~al.}(1992)\textit{Shibata, Ishido, Acton, Strong,
  Hirayama, Uchida, McAllister, Matsumoto, Tsuneta, Shimizu, Hara, Sakurai,
  Ichimoto, Nishino, and Ogawara}}]{Setal92}
Shibata, K., Y.~Ishido, L.~W. Acton, K.~T. Strong, T.~Hirayama, Y.~Uchida,
  A.~H. McAllister, R.~Matsumoto, S.~Tsuneta, T.~Shimizu, H.~Hara, T.~Sakurai,
  K.~Ichimoto, Y.~Nishino, and Y.~Ogawara (1992), Observations of x-ray jets
  with the yohkoh soft x-ray telescope, \textit{\pasj}, \textit{44},
  L173--L179.

\bibitem[{\textit{{Shibata} et~al.}(2013)\textit{{Shibata}, {Isobe}, {Hillier},
  {Choudhuri}, {Maehara}, {Ishii}, {Shibayama}, {Notsu}, {Notsu}, {Nagao},
  {Honda}, and {Nogami}}}]{Setal13}
{Shibata}, K., H.~{Isobe}, A.~{Hillier}, A.~R. {Choudhuri}, H.~{Maehara}, T.~T.
  {Ishii}, T.~{Shibayama}, S.~{Notsu}, Y.~{Notsu}, T.~{Nagao}, S.~{Honda}, and
  D.~{Nogami} (2013), {Can Superflares Occur on Our Sun?}, \textit{\pasj},
  \textit{65}, 49, \doi{10.1093/pasj/65.3.49}.

\bibitem[{\textit{{Somov}}(1992)}]{S92}
{Somov}, B.~V. (Ed.) (1992), \textit{{Physical processes in solar flares.}},
  \textit{Astrophysics and Space Science Library}, vol. 172, Springer,
  Netherlands, \doi{10.1007/978-94-011-2396-9}.

\bibitem[{\textit{{Spicer}}(1982)}]{S82}
{Spicer}, D.~S. (1982), {Magnetic energy storage and conversion in the solar
  atmosphere}, \textit{\ssr}, \textit{31}, 351--435, \doi{10.1007/BF00171370}.

\bibitem[{\textit{{Spicer} and {Sudan}}(1984)}]{SS84}
{Spicer}, D.~S., and R.~N. {Sudan} (1984), {Beam-return current systems in
  solar flares}, \textit{\apj}, \textit{280}, 448--456, \doi{10.1086/162011}.

\bibitem[{\textit{{Sturrock}}(1966)}]{S66}
{Sturrock}, P.~A. (1966), {Model of the High-Energy Phase of Solar Flares},
  \textit{\nat}, \textit{211}, 695--697, \doi{10.1038/211695a0}.

\bibitem[{\textit{{Svestka}}(1976)}]{Sv76}
{Svestka}, Z. (1976), \textit{{Solar Flares}}, 415 pp., Springer-Verlag Berlin
  Heidelberg 1976.

\bibitem[{\textit{{Swann}}(1933)}]{S33}
{Swann}, W.~F. (1933), {A Mechanism of Acquirement of Cosmic-Ray Energies by
  Electrons}, \textit{\pr}, \textit{43}, 217--220,
  \doi{10.1103/PhysRev.43.217}.

\bibitem[{\textit{{Sweet}}(1958)}]{S58}
{Sweet}, P.~A. (1958), {The Neutral Point Theory of Solar Flares}, in
  \textit{Electromagnetic Phenomena in Cosmical Physics}, \textit{IAU
  Symposium}, vol.~6, edited by B.~{Lehnert}, p. 123.

\bibitem[{\textit{{Tandberg-Hanssen} and {Emslie}}(1988)}]{T-HE88}
{Tandberg-Hanssen}, E., and A.~G. {Emslie} (1988), \textit{{The physics of
  solar flares}}, Cambridge University Press.

\bibitem[{\textit{{T{\"o}r{\"o}k} and {Kliem}}(2003)}]{TK03}
{T{\"o}r{\"o}k}, T., and B.~{Kliem} (2003), {The evolution of twisting coronal
  magnetic flux tubes}, \textit{\aap}, \textit{406}, 1043--1059,
  \doi{10.1051/0004-6361:20030692}.
 

\bibitem[{\textit{{T{\"o}r{\"o}k} et~al.}(2004)\textit{{T{\"o}r{\"o}k},
  {Kliem}, and {Titov}}}]{TKT04}
{T{\"o}r{\"o}k}, T., B.~{Kliem}, and V.~S. {Titov} (2004), {Ideal kink
  instability of a magnetic loop equilibrium}, \textit{\aap}, \textit{413},
  L27--L30, \doi{10.1051/0004-6361:20031691}.

\bibitem[{\textit{{T{\"o}r{\"o}k} et~al.}(2014)\textit{{T{\"o}r{\"o}k},
  {Leake}, {Titov}, {Archontis}, {Miki{\'c}}, {Linton}, {Dalmasse}, {Aulanier},
  and {Kliem}}}]{Tetal14}
{T{\"o}r{\"o}k}, T., J.~E. {Leake}, V.~S. {Titov}, V.~{Archontis},
  Z.~{Miki{\'c}}, M.~G. {Linton}, K.~{Dalmasse}, G.~{Aulanier}, and B.~{Kliem}
  (2014), {Distribution of Electric Currents in Solar Active Regions},
  \textit{\apjl}, \textit{782}, L10, \doi{10.1088/2041-8205/782/1/L10}.

\bibitem[{\textit{{van den Oord}}(1990)}]{vdO90}
{van den Oord}, G.~H.~J. (1990), {The electrodynamics of beam/return current
  systems in the solar corona}, \textit{\aap}, \textit{234}, 496--518.

\bibitem[{\textit{{van Driel-Gesztelyi} et~al.}(2003)\textit{{van
  Driel-Gesztelyi}, {D{\'e}moulin}, and {Mandrini}}}]{vDGetal03}
{van Driel-Gesztelyi}, L., P.~{D{\'e}moulin}, and C.~H. {Mandrini} (2003),
  {Observations of magnetic helicity}, \textit{Advances in Space Research},
  \textit{32}, 1855--1866, \doi{10.1016/S0273-1177(03)90619-3}.

\bibitem[{\textit{{van Tend} and {Kuperus}}(1978)}]{vTK78}
{van Tend}, W., and M.~{Kuperus} (1978), {The development of coronal electric
  current systems in active regions and their relation to filaments and
  flares}, \textit{\solphys}, \textit{59}, 115--127, \doi{10.1007/BF00154935}.

\bibitem[{\textit{{Vasyliunas}}(2005)}]{V05}
{Vasyliunas}, V.~M. (2005), {Relation between magnetic fields and electric
  currents in plasmas}, \textit{Annales Geophysicae}, \textit{23}, 2589--2597,
  \doi{10.5194/angeo-23-2589-2005}.

\bibitem[{\textit{{Wheatland} and {Melrose}}(1994)}]{WM94}
{Wheatland}, M.~S., and D.~B. {Melrose} (1994), {Cross-field current closure
  below the solar photosphere}, \textit{\ajp}, \textit{47}, 361--374,
  \doi{10.1071/PH940361}.

\bibitem[{\textit{{Wiegelmann} and {Sakurai}}(2012)}]{WS12}
{Wiegelmann}, T., and T.~{Sakurai} (2012), {Solar Force-free Magnetic Fields},
  \textit{Living Reviews in Solar Physics}, \textit{9}, 5,
  \doi{10.12942/lrsp-2012-5}.

\bibitem[{\textit{{Woods} and {Thompson}}(2006)}]{WT06}
{Woods}, P.~M., and C.~{Thompson} (2006), {Soft gamma repeaters and anomalous
  X-ray pulsars: magnetar candidates}, in \textit{Compact stellar X-ray
  sources}, edited by W.~H.~G. {Lewin} and M.~{van der Klis}, pp. 547--586,
  Cambridge University Press.

\bibitem[{\textit{{Zhang} et~al.}(2006)\textit{{Zhang}, {Flyer}, and
  {Low}}}]{ZFL06}
{Zhang}, M., N.~{Flyer}, and B.~C. {Low} (2006), {Magnetic Field Confinement in
  the Corona: The Role of Magnetic Helicity Accumulation}, \textit{\apj},
  \textit{644}, 575--586, \doi{10.1086/503353}.

\bibitem[{\textit{{Zuccarello} et~al.}(2014)\textit{{Zuccarello}, {Seaton},
  {Mierla}, {Poedts}, {Rachmeler}, {Romano}, and {Zuccarello}}}]{Zetal14}
{Zuccarello}, F.~P., D.~B. {Seaton}, M.~{Mierla}, S.~{Poedts}, L.~A.
  {Rachmeler}, P.~{Romano}, and F.~{Zuccarello} (2014), {Observational Evidence
  of Torus Instability as Trigger Mechanism for Coronal Mass Ejections: The
  2011 August 4 Filament Eruption}, \textit{\apj}, \textit{785}, 88,
  \doi{10.1088/0004-637X/785/2/88}.

\bibitem[{\textit{{Zuccarello} et~al.}(2015)\textit{{Zuccarello}, {Aulanier},
  and {Gilchrist}}}]{Zetal15}
{Zuccarello}, F.~P., G.~{Aulanier}, and S.~A. {Gilchrist} (2015), {Critical
  Decay Index at the Onset of Solar Eruptions}, \textit{\apj}, \textit{814},
  126, \doi{10.1088/0004-637X/814/2/126}.

\end{thebibliography}

\listofchanges
Changes are denoted in bold magenta font is a separately uploaded PDF.

\end{document}